\documentclass[a4paper,12pt]{article}
\usepackage{amsbsy}
\usepackage[fleqn]{amsmath}
\usepackage{amssymb}
\usepackage{array}
\usepackage{booktabs}
\usepackage{xcolor}
\usepackage[colorlinks, linkcolor=blue, anchorcolor=blue, citecolor=blue]{hyperref}
\usepackage{mathrsfs}
\usepackage{tabularx}
\usepackage{latexsym}
\usepackage{textcomp}
\usepackage{pstricks}
\usepackage{colortbl}
\usepackage{multirow}
\usepackage{fancyhdr}
\usepackage{bm}
\usepackage{booktabs}
\usepackage{subfigure}
\usepackage{caption}
\usepackage{soul}
\usepackage{cleveref}
\usepackage{cite}
\usepackage{lineno}
\usepackage{ulem}
\usepackage{graphicx}
\usepackage{epstopdf}
\usepackage{indentfirst}
\usepackage{setspace}
\usepackage[latin1]{inputenc}
\usepackage{titlesec}
\usepackage{tabularx} 
\usepackage{ragged2e} 
\usepackage{booktabs} 
\usepackage{float}  
\usepackage{stmaryrd}
\usepackage{amsthm}
\usepackage{amssymb}
\SetSymbolFont{stmry}{bold}{U}{stmry}{m}{n}

\titleformat*{\section}{\large\bf}
\titleformat*{\subsection}{\normalsize\bf}
\crefname{section}{Section}{Sections}
\crefname{equation}{Eq.}{Eqs.}
\crefname{figure}{Figure}{Figs.}
\crefname{table}{Table}{Tables}
\newcommand{\gv}[1]{\ensuremath{\mbox{\boldmath$ #1 $}}}
\newcommand{\abs}[1]{\left| #1 \right|} 
\newcommand{\pd}[2]{\frac{\partial #1}{\partial #2}}
\newcommand{\grad}[1]{\gv{\nabla} #1} 
\renewcommand{\div}[1]{\gv{\nabla} \cdot #1} 

\newcommand{\RomanOne}{\mathbf{\uppercase\expandafter{\romannumeral1}}}
\newcommand{\RomanTwo}{\mathbf{\uppercase\expandafter{\romannumeral2}}}
\begin{document}

%
%
\begin{center}
\Large\textbf{DEDEM: Discontinuity Embedded Deep Energy Method for solving fracture mechanics problems}
\end{center}
\large{
\begin{center}
\textbf{Luyang Zhao}$^{a}$, \textbf{Qian Shao}$^{a,b}$\footnote{Corresponding author. E-mail address: qian.shao@whu.edu.cn.}$^{}$
\end{center}
}
\small{
\begin{center}
$^a$Department of Engineering Mechanics, School of Civil Engineering, Wuhan University, 430072 Wuhan, China\\
$^b$Wuhan University Shenzhen Research Institute, 518057 Shenzhen, China\\
\end{center}
}

%
%
\begin{flushleft}
\large\textbf{Abstract}
\end{flushleft}
\indent\indent 
Physics-Informed Neural Networks (PINNs) have aroused great attention for its ability to address forward and inverse problems of partial differential equations. However, approximating discontinuous functions by neural networks poses a considerable challenge, which results in high computational demands and low accuracy to solve fracture mechanics problems within standard PINNs framework. In this paper, we present a novel method called Discontinuity Embedded Deep Energy Method (DEDEM) for modeling fracture mechanics problems. In this method, interfaces and internal boundaries with weak/strong discontinuities are represented by discontinuous functions constructed by signed distance functions, then the representations are embedded to the input of the neural network so that specific discontinuous features can be imposed to the neural network solution. Results demonstrate that DEDEM can accurately model the mechanical behaviors of cracks on a large variety of fracture problems. Besides, it is also found that DEDEM achieves significantly higher computational efficiency and accuracy than the existing methods based on domain decomposition techniques. 
\begin{flushleft}
\end{flushleft}

%
%
\section{Introduction}\label{introduction}

Fracture mechanics analysis is of utmost importance for prediction of failure mode and assessment of service life in engineering materials and structures. Numerical simulations of the fracture behaviour involve solving partial differential equations (PDEs) with singularities at the crack tip and jump discontinuities across the crack surface, which pose significant challenges for numerical modeling, especially for the classic finite element method (FEM) \cite{zienkiewicz2005finite} which is the most popular tool for solid mechanics modeling. Over the past decades, various numerical tools have been developed to model the mechanical behaviour of the crack, including the Extended Finite Element Method (XFEM) \cite{Moes1999131}, boundary Element Method (BEM) \cite{Aliabadi199783}, phase field method (PF) \cite{francfort1998revisiting} and peridynamics \cite{silling2007peridynamic}.


The advent of increased computational power has led to a resurgence in the use of neural networks (NNs) for solving PDEs, decades after the initial pioneering work \cite{lee1990neural}. Thanks to the universal approximator property of NNs \cite{hornik1989multilayer}, the Physics-Informed Neural Networks (PINNs) framework proposed by Rassi et al. \cite{raissi2019physics} is able to deal with forward and inverse problems involving PDEs by employing NNs as function approximators. Subsequently, PINNs have found success in a variety of PDE-governed physical systems, including fluid mechanics \cite{cai2021fluid}, heat transfer problems \cite{cai2021heat} and cardiovascular flows \cite{kissas2020machine}. Additionally, Haghighat et al. \cite{haghighat2021physics} extended the application of PINNs to solving forward problems, inversion and surrogate modeling for solid mechanics problems. The standard PINNs framework involves enforcing PDEs and initial/boundary conditions by minimizing squared errors on collocation points inside the domain and on the boundaries, respectively. An alternative training strategy is the Deep Ritz Method (DRM) \cite{yu2018deep}, which focuses on addressing the variational formulations of PDEs with NNs. DRM can more efficiently solve the PDEs as the method requires lower orders of gradients and automatically satisfies Neumann boundary conditions. Subsequently, DRM is introduced to solving solid mechanics problems, and the implementation is called deep energy method (DEM) \cite{samaniego2020energy}, whose application involves elastic plate \cite{li2021physics}, hyperelasticity \cite{fuhg2022mixed} and topology optimization \cite{jeong2023complete}.

Nonetheless, the application of PINNs approach encounters formidable challenges in tackling problems involving strong and weak discontinuities as feed-forward networks have limited ability on discontinuous functions \cite{llanas2008constructive}. The difficulties also present challenges in solving fracture problems. Many attempts to modeling fracture avoid directly describing the discontinuities. Some researchers utilize the phase field model \cite{samaniego2020energy,goswami2020transfer,manav2024phase,zheng2022physics}, where cracks are viewed as material damage and dispersed to be process zones with finite width; Other researchers utilize peridynamics theories \cite{yu2023nonlocal,ning2023physics}, where cracking behaviors are modeled by non-local interactions. Despite success on modeling crack behavior achieved by these methods, they cannot directly deal with ideal line cracks. The question of modeling strong and weak discontinuities problems with PINNs still not be answered.

One straightforward approach to overcome such challenge is to use domain decomposition techniques, where the entire domain is split into several sub-domains by the interfaces and each subdomain is approximated using an individual neural network. The initial attempt at domain decomposition was made by conservative-PINNs (CPINNs) \cite{jagtap2020conservative}, followed by the subsequent work extended-PINNs (XPINNs) \cite{jagtap2021extended} proposed by Rassi et al. Then Wu et al. \cite{wu2022inn} proposed interfaced neural networks (INN) to improve the efficiency and accuracy for domain decomposition techniques. He et al. \cite{he2022mesh} applied domain decomposition to solve 3D elliptic interface problems. Yu et al. \cite{diao2023solving} used CPINNs to solve multi-material problems in solid mechanics. Jang et al. \cite{jang2024partitioned} applied augmented Lagrangian term to accelerate the convergence for solving interface problems. Domain decomposition techniques are then be employed to solve fracture mechanics problems. Wang et al. \cite{wang2022cenn} proposed conservative energy method (CENN) to employ domain decomposition within DEM framework, which successfully simulated mode III crack and heterogeneous problems. Gu et al. \cite{gu2023enriched} incorporated enrichment functions to precisely capture the singular crack-tip field and calculate the stress intensity factors (SIFs) of 2D in-plane cracks. Then the same technique is utilized to solve interface fracture problems \cite{gu2024interface} and fatigue crack growth \cite{chen2024crack}.

However, several challenges remain in domain decomposition based method: 

(1) $\textit{Intractable hyperparameter tuning:}$ The interface conditions between sub-domains are enforced through additional penalties in the loss function. It has been proven that the hyperparameters associated with these loss terms are crucial for the successful training of both strong and weak forms \cite{jagtap2020conservative,wang2022cenn}. Balancing these hyperparameters becomes particularly challenging for too many loss terms introduced by domain decomposition. 

(2) $\textit{High computational cost}$: Training separate NNs for each sub-domain results in an increased number of trainable parameters. Without the implementation of parallelization techniques \cite{shukla2021parallel}, this computational burden will reach an impractical level for training sub-networks in serial. Besides, competition among multiple loss terms also significantly impacts the training efficiency \cite{wang2022NTK}, as much more training epochs are required to achieve convergence.

(3) $\textit{High risk for training failure}$: The competition among multiple loss terms mentioned above also significantly impacts the accuracy. The imbalanced multi-objective optimisation process sometimes leads to training failures caused by gradient pathologies \cite{wang2021understanding}. 

To overcome these challenges, efforts has been made to modeling discontinuities bypassing the domain decomposition techniques to improve the efficiency and accuracy. Wang et al. \cite{wang2020mesh} first investigated DRM on solving interface problems with discontinuous coefficients. Hu et al. \cite{hu2022discontinuity} proposed DCSNN to approximate piece-wise continuous function by introducing additional feature inputs to the network. Subsequently, Tseng et al. \cite{tseng2023cusp} enable the neural network to capture weak discontinuities for elliptic interface problems using the same approach. Yao et al. \cite{yao2023deep} solved multi-material diffusion equation with one single network by domain separation strategy, and the discontinuities are provided by the separated inputs to the network. Sarma et al. \cite{sarma2024interface} proposed interface PINNs (I-PINNs) to model interface problems with modified NN architecture. It should be noted that there still lacks investigation on modeling internal boundaries, so that problems with ideal line crack cannot be solved.

In this paper, we propose the Discontinuity Embedded Deep Energy Method (DEDEM), which is applied to obtain weak solutions for fracture mechanics problems. Within our framework, interfaces and cracks are tracked by signed distance functions (SDF), then strong and weak discontinuities are imposed to the neural network solution through embedding the discontinuities into the inputs. Our method offers the following advantages:

(1)	$\textit{Flexibility}$: Our approach for constructing discontinuities can be readily applied to model arbitrary crack patterns, including complex curved cracks, crossing cracks and bi-material interface cracks. Besides, such description by SDF is easy for implementation.

(2)	$\textit{Efficiency and Accuracy}$: Without competition between interface conditions and governing equations during the training process, 
the neural network converges much faster and higher accuracy can be achieved. Besides, crack-tip singularities can be automatically captured by NNs without any special treatment.

(3)	$\textit{Convenience for postprocessing}$: PINNs provide a meshless solution in the whole domain, which offer convenience for any subsequent calculation, including acquiring stress intensity factors and computing the direction for crack propagation.

The paper is organized as follows. \cref{Methodology} introduces the governing equations for discontinuous elastic problems and basic idea of PINNs method. \cref{Methodology} details the proposed DEDEM for modeling problems with strong/weak discontinuities. Then the implementation details and four examples are presented in \cref{Results} to show the applicability of DEDEM on various fracture problems. Subsequently, \cref{Discussion} discussed some possible extensions of our proposed method. Finally, a brief summary is given in \cref{conclusion}.

%
%
\section{Preliminaries}\label{Preliminaries}

    \subsection{Governing equations for discontinuous elasticity problems}\label{elasticity}

    The governing equation of a quasi-static mechanical system is given by the equilibrium equations under prescribed boundary conditions:

    \begin{equation}
        \begin{aligned}
            \div{\bm{\sigma}}+\bm{b}&=0 \text{ in }\Omega , \\
            \bm{u}&= \bm{\bar{u}} \text{ on }\Gamma_u ,\\
            \bm{\sigma} \cdot \bm{n}&=\bm{\bar{t}}\text{ on }\Gamma_t ,
            \label{equilibrium}
        \end{aligned}
    \end{equation}
    where $\bm{\sigma}$ is the stress tensor, $ \bm{b}$ is the body force,
    $ \bm{u}$ is the displacement vector, $\bm{n}$ is the unit outward normal, 
    $\bm{\bar{u}}$ and  $\bm{\bar{t}}$ respectively denotes for the displacement and force boundary conditions; $\Omega$ is the computational domain and $\Gamma_u$ and $\Gamma_t$ are respectively essential and natural boundaries.
    Then the constitutive relation of linear elastic problems can be formulated by:
    \begin{equation}
        \bm{\sigma}=\bm{C} : \bm{\varepsilon} ,\\
        \label{constitutive}
    \end{equation}
    where $ \bm{C}$ is the stiffness tensor which can be defined by the Young's modulus $ E$ 
    and the Poisson's ratio $\nu$ for linear elastic problems. $\bm{\varepsilon}$ is the strain tensor:
    \begin{equation}
        \bm{\varepsilon}=\grad_s{\bm{u}} \text{ with } \varepsilon_{ij}=\frac{1}{2} \left(\pd{u_i}{x_j}+\pd{u_j}{x_i}\right) ,
        \label{geometry}
    \end{equation}  
    where $\grad_s$ is the symmetric part of the gradient operator. 

\begin{figure}[!hbtp]
        \centering
        \includegraphics[]{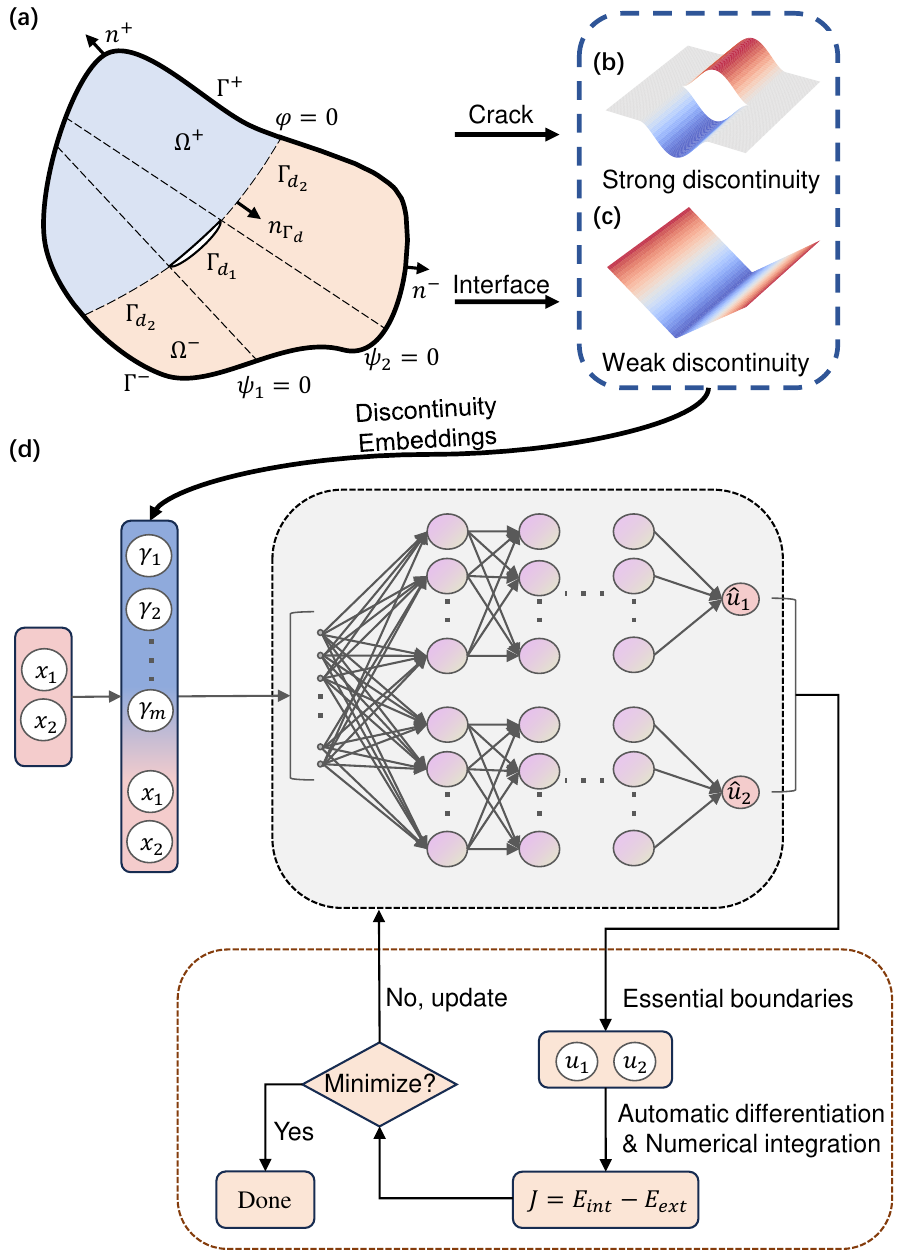}
        \caption{The schematic of DEDEM: (a) The schematic of a problem with a crack; (b) strong and (c) weak discontinuous functions constructed from SDF; (d) The whole training workflow.
        }
        \label{overview}
    \end{figure}

    In order to derive the governing equation for discontinuous elasticity problems, here we consider a discontinuous problem shown in \cref{overview} (a). The whole domain $\Omega$ is divided into $\Omega^{+}$ and $\Omega^{-}$ by the interface $\Gamma_{d}$, and the external boundaries, where $\Gamma_{d}=\Gamma_{d_1}\cup \Gamma_{d_2}$. Here both of the strong and weak discontinuities exist, as the solution is discontinuous across the crack surface $\Gamma_{d_1}$, and the normal derivative is discontinuous across the material interface $\Gamma_{d_2}$. 
    The interface conditions on $\Gamma_{d}$ can be formulated as:
    \begin{equation}
        \begin{aligned}
            \bm{\sigma}^{+} \cdot \bm{n}^{+} +  \bm{\sigma}^{-} \cdot \bm{n}^{-} &= \llbracket{\bm{\bar{t_{d}}}}\rrbracket \text{ on }\Gamma_{d_1} ,\\
            \bm{\sigma}^{+} \cdot \bm{n}^{+} + \bm{\sigma}^{-} \cdot \bm{n}^{-} &=  0 \text{ on }\Gamma_{d_2} ,\\
            \bm{u}^{+} - \bm{u}^{-} &=  0\text{ on }\Gamma_{d_2} ,
            \label{crack_condition}
        \end{aligned}
    \end{equation}    
    where $\llbracket{\bm{\bar{t_{d}}}}\rrbracket$ denotes the traction acting on the crack surface. Here we let $\bm{n_{\Gamma_{d}}}=\bm{n}^{+}=-\bm{n}^{-}$ on $\Gamma_{d}$.
    Since the differential operator of \cref{equilibrium} is linear and self-adjoint \cite{zienkiewicz2005finite}, then the weak formulation can be derived by introducing a trial function $\bm{u}$ and a test function $\delta\bm{u}$:
    \begin{equation}
        \int_{\Omega} \delta\bm{u} \cdot (\div{\bm{\sigma}(\bm{u})}+\bm{b}) d\Omega = 0 ,
        \label{virtual_work}
    \end{equation}  
    where both of the test and trial function must satisfy the essential boundary conditions.
    Then integration by parts is used:
    \begin{equation}
        \int_{\Omega} \delta(\grad{\bm{u}}) : {\bm{\sigma}(\bm{u})} d\Omega
        - \int_{\Omega} \div{(\delta \bm{u} \cdot \bm{\sigma}(\bm{u}))} d\Omega
        - \int_{\Omega} \delta\bm{u} \cdot \bm{b} d\Omega
        = 0  .
        \label{int_by_parts}
    \end{equation}  
    The divergence theorem is applied  respectively in $\Omega^{+}$ and $\Omega^{-}$ and 
    interface conditions are utilized, the second term of \cref{int_by_parts} can be formulated as:
    \begin{equation}
        \begin{aligned}
        \int_{\Omega} \div{(\delta \bm{u} \cdot \bm{\sigma})} d\Omega &= 
            \int_{\Omega^{+}} \div{(\delta \bm{u} \cdot \bm{\sigma})} d\Omega + 
            \int_{\Omega^{-}} \div{(\delta \bm{u} \cdot \bm{\sigma})} d\Omega \\
        &= \int_{\Gamma^{+}\cup\Gamma_{d1}\cup \Gamma_{d2}} \delta \bm{u}^{+} \cdot \bm{\sigma}^{+} \cdot \bm{n^{+}} d\Gamma +
            \int_{\Gamma^{-}\cup\Gamma_{d1}\cup \Gamma_{d2}} \delta \bm{u}^{-} \cdot \bm{\sigma}^{-} \cdot \bm{n^{-}} d\Gamma \\
        &= \int_{\Gamma } \delta \bm{u} \cdot \bm{\sigma} \cdot \bm{n} d\Gamma + 
            \int_{\Gamma_{d1}\cup \Gamma_{d2}} \delta \bm{u} \cdot (\bm{\sigma}^{+} \cdot \bm{n}^{+} +  \bm{\sigma}^{-} \cdot \bm{n}^{-}) d\Gamma \\
        &=  \int_{\Gamma } \delta \bm{u} \cdot \bm{\bar{t}} d\Gamma + 
            \int_{\Gamma_{d_2}} \delta \bm{u} \cdot \llbracket{\bm{\bar{t_{d}}}}\rrbracket d\Gamma .
            \label{div_discontinuous}
        \end{aligned}
    \end{equation} 
    By utilizing the symmetry of stress tensor on the first term and \cref{div_discontinuous} on the second term, the weak form of governing equation for the discontinuities problem can be written as:
    \begin{equation}
        \delta J(\bm{u})
        = \int_{\Omega} \delta\bm{\varepsilon}(\bm{u}) : \bm{C} : \bm{\varepsilon}(\bm{u}) d\Omega
        - \int_{\Gamma_t} \delta\bm{u} \cdot \bm{\bar{t}} d\Gamma
        - \int_{\Gamma_{d_2}} \delta\bm{u} \cdot \llbracket{\bm{\bar{t_{d}}}}\rrbracket d\Gamma
        - \int_{\Omega} \delta\bm{u} \cdot \bm{b} d\Omega
        = 0 ,
        \label{discontinuous_weak_form}
    \end{equation}
    where $J(\bm{u})$ is the total potential energy:   
    \begin{equation}
        J(\bm{u})
        = \int_{\Omega} \frac{1}{2} \bm{\varepsilon}(\bm{u}) : \bm{C} : \bm{\varepsilon}(\bm{u}) d\Omega
        - \int_{\Gamma_t} \bm{u} \cdot \bm{\bar{t}} d\Gamma  
        - \int_{\Gamma_{d_2}} \bm{u} \llbracket{\bm{\bar{t_{d}}}}\rrbracket d\Gamma  
        - \int_{\Omega} \bm{u} \cdot \bm{b} d\Omega.
        \label{potential_energy}
    \end{equation}
    The first term stands for the strain energy, and the rest terms respectively stands for 
    the external energy contributed by surface traction on external boundaries and internal boundaries, and the body force inside domain.
    It should be noted that the formulation of the principle of virtual work is equivalent to minimize $J(\bm{u})$.

    \subsection{Physics-informed Neural Networks} \label{PINN}

    In Physics-Informed Neural Networks (PINNs) framework, solutions of PDEs are approximated by deep learning techniques. 
    Deep Neural Networks (DNNs), which have been proven to be universal approximators \cite{hornik1989multilayer}, are the primary choice for PINNs. 
    A DNN can be formulated as a composition of simple transformations:
    \begin{equation}
        y = NN(x) = T^{(n)}(T^{(n-1)}(...T^{(1)}(x))),
        \label{DNN}
    \end{equation}
    where $x$ is the input, and $y$ is the output. 
    $T$ is one single hidden layer, which is a combination of a linear transformation and an activation function:
    \begin{equation}
        y^{(i)} = T^{(i)}(x^{(i)}) = \sigma(W^{(i)}x^{(i-1)} + b^{(i)}),
        \label{single_layer}
    \end{equation}
    where $x^{(i)}$ and $y^{(i)}$ are the inputs and outputs of the $i-th$ layer, respectively, 
    while $W$ and $b$ are trainable parameters that are updated by gradient descent and back propagation algorithms. 
    The activation function introduces nonlinearity to the neural network. 
    Then we consider a PDE with given boundary conditions on $\partial{\Omega}:=\cup_{n = 1}^{M}\Gamma_i$:
    \begin{equation}
        \begin{aligned}
        \bm{\mathcal{L} }(\bm{u}) &= f \text{ in }\Omega, \\
        \bm{\mathcal{G} }(\bm{u}) &= g_i \text{ on }\Gamma_i \text{ }(i=1,2,3,...,M),
        \label{PDE}
        \end{aligned}
    \end{equation}
    where $\bm{\mathcal{L} }$ is a known differential operator, and $\bm{\mathcal{G} }$ denotes arbitrary boundaries conditions, including Dirichlet and Neumann boundaries. 
    Let $\bm{u}$ be the approximation of the neural networks,
    the solution of the PDE can be approximated by the neural network by minimizing the following loss function:
    \begin{equation}
        \bm{u}(\bm{x};\theta) = \underset{\theta}{\arg \min} L(\theta)
        := \underbrace{\frac{1}{N_r}\sum_{i = 1}^{N_r} {\left\lVert \bm{\mathcal{L}}\bm{u\left(x^{(i)}\right)} - f \right\rVert}^2 }_{\text{PDE loss}}  
        + \underbrace{\sum_{j = 1}^{M}{\frac{1}{N_b^{(j)}}\sum_{i = 1}^{N_b^{(j)}} {\left\lVert \bm{\mathcal{G}}\bm{u\left(x^{(i)}\right)} - g_j \right\rVert}^2 }}_{\text{B.C. loss}}  ,
        \label{PINN_loss}
    \end{equation}
    where $\theta$ is the learnable parameters. If \cref{PDE} has a variational form, the solution can also be obtained by:
    \begin{equation}
        \bm{u}(\bm{x};\theta) = \underset{\theta}{\arg \min} I(\theta)
        := J(\bm{u})
        + \beta \int_{\Gamma_u} {\left\lVert (\bm{u} - \bar{\bm{u}} \right\rVert}^2  \,d\Gamma_u ,
        \label{DEM_loss}
    \end{equation}
    where $\Gamma_u$ is the boundaries with essential boundary conditions $\bar{\bm{u}}$, and $\beta$ is the lagrangian multiplier, which is a hyper-parameter needs to be tuned. 
    Using the energy form of PDE for training offers two advantages over strong form: 
    1.Training with energy form requires lower orders of derivative than strong form, leading to a substantial reduction in computational cost during the back propagation training process.
    2.Neumann boundary conditions can be automatically satisfied, so that less loss terms and hyperparameters are required. 
    By substituting $J(\bm{u(\bm{x};\theta)})$ with \cref{potential_energy}, linear elasticity problems can be solved.
    In this paper, we choose to use the weak form to train the network. 
    The NN approximation is trained by the loss function composed of potential energy and essential boundary conditions.
    It should be noted that the traction-free conditions can be satisfied without specific treatment, which offers convenience in implementation.

%
%
\section{Discontinuity Embedded Deep Energy Method}\label{Methodology}
\subsection{Overview of the proposed method}
In standard PINNs, the solution is approximated by a neural network, where the inputs are the spatial coordinates of the sampling points. It has been proven theoretically \cite{llanas2008constructive} and experimentally \cite{sarma2024interface} that the output solutions can hardly output discontinuous solutions. This makes it challenging to approximate discontinuous physical fields with PINNs. In DEDEM, we introduce a transformation of spatial coordinates and embed them into the input of the neural network:
\begin{equation}
    \bm{u}(\bm{x}) = u_{NN}(\bm{x}, \bm{\gamma}(\bm{x})),
    \label{augumented_axis}
\end{equation}
where $u_{NN}$ represents the neural network, $\bm{x}$ is the spatial coordinates of points, and $\bm{\gamma} = \{\gamma_1, \gamma_2, ..., \gamma_n\}$ denotes the introduced embedding to describe the discontinuities inside the domain. The derivative of the output respect to a spatial coordinate $x_i$ can be given by the chain rule:
\begin{equation}
    \frac{\partial \bm{u}(\bm{x})} {\partial x_i} = 
    \frac{\partial u_{NN}(\bm{x}, \bm{\gamma}(\bm{x}))}{\partial x_i} 
    + \frac{\partial u_{NN}(\bm{x}, \bm{\gamma}(\bm{x}))}{\partial \gamma(\bm{x})} \cdot \frac{\partial\gamma(\bm{x})}{\partial x_i} .
    \label{augumented_axis_derivative}
\end{equation}
It is obvious that if the embedding $\bm{\gamma}(\bm{x})$ and the derivative of embedding $\frac{\partial\gamma(\bm{x})}{\partial x_i}$ is constructed to satisfy specific features, the outputs of NN can be imposed to satisfy specific discontinuities. 
In this section, we provide a detailed description of the construction of embeddings for describing strong and weak discontinuities across the interfaces.

\subsection{Strong discontinuities on opening interfaces}

\begin{figure}[!ht]
    \centering
    \includegraphics[]{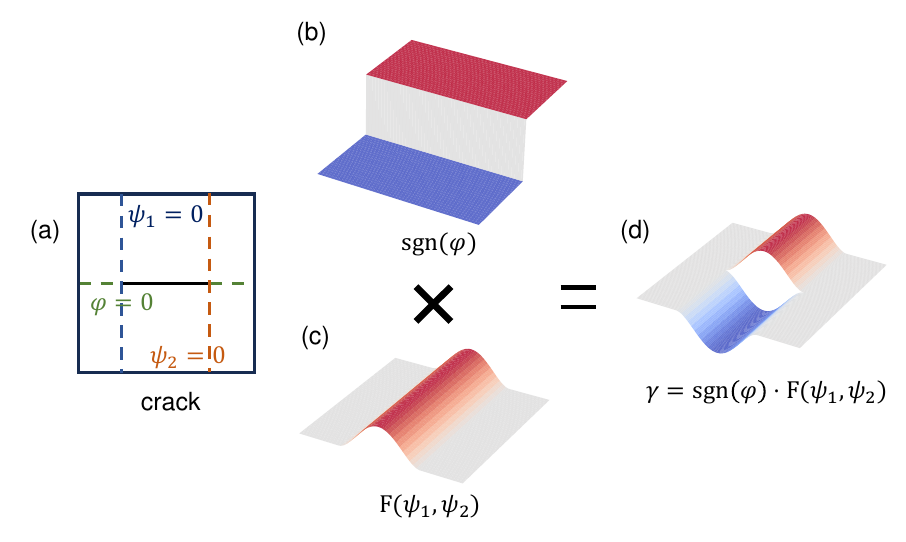}
    \caption{An example of the embedding of a crack: (a) A domain with a crack; (b) the sign function constructed from $\varphi$ to separate the crack surface and (c) continuous functions constructed from $\psi$ to guarantee the continuity outside the crack; (d) The constructed embedding $\gamma$ for describing the crack.
    }
    \label{fig_strong_discontinuity}
\end{figure}

Strong discontinuities occur in crack problems as displacements are discontinuous across the crack surface. 
Displacements on one side are distinct from the other side on the crack surface, leading to multiple values correspond to one spatial coordinate. 
The displacement approximation must have non-zero solution jumps across the crack, 
which can be satisfied when $\gamma$ is selected to be $\gamma\left(\bm{x}^+\right)\neq\gamma\left(\bm{x}^-\right)$:
\begin{equation}
    \llbracket{\bm{u}}\rrbracket = \bm{u}^+ - \bm{u}^- =
    NN(\bm{x},\gamma^+(\bm{x})) -  NN(\bm{x},
    \gamma^-(\bm{x})) \neq 0 .
    \text{ on crack surface}
    \label{crack_embed}
\end{equation}
This can be simply achieved by selecting $\gamma$ as the sign function:
\begin{equation}
    \gamma = \operatorname{sgn}{(\varphi)} ,
    \label{sgn_func}
\end{equation}
where $\operatorname{sgn}{(\cdot)}$ takes the following form:
\begin{equation}
    \operatorname{sgn}{(x)}=
    \left\{ 
        \begin{array}{lc}
            1 & x > 0 \\
            -1 & x \leq 0\\
        \end{array}
    \right. ,
    \label{sgn}
\end{equation}
and $\varphi$ is the signed distance functions (SDF) to track and describe the crack surface. SDF techniques is widely used for tracking and describing surfaces and interfaces in XFEM \cite{belytschko2001arbitrary}. An example of an interior crack is given in \cref{fig_strong_discontinuity} (a), and the described surface is the green line, of which the $\varphi$ can be given by: 
\begin{equation}        
    \varphi(\bm{x}) = \operatorname{sgn}{\left( \bm{n} \cdot (\bm{x} - \bm{\hat{x}}) \right)} ||\bm{x} - \bm{\hat{x}}|| ,
    \label{SDF}
\end{equation}
where $\bm{\hat{x}}$ denotes for the closest point on the surface to the point $\bm{x}$, $\bm{n}$ is the outward normal, and $||\bm{x} - \bm{\hat{x}}||$ is the distance of the point $\bm{x}$ to the surface.

Since the crack do not divide the domain into two distinct parts completely, 
the sign function needs to be modified to guarantee the smoothness in front of the crack tip. 
This can be achieved by modifying the embedding with SDF for describing the tangential extension of the crack tip. 
For an interior crack, the embedding can be given by:
\begin{equation}
    \gamma = \operatorname{F}\left(\psi_1,\psi_2\right) \cdot \operatorname{sgn}(\varphi)
    \label{interior_crack_1} ,
\end{equation}
where $\psi_1,\psi_2$ are the SDF of the crack tips on tangential direction, and $\operatorname{F}\left(\psi_1,\psi_2\right)$ requires to have C1 continuity to ensure the flux continuities in front of the crack tip. The selection of $\operatorname{F}(\cdot)$ is arbitrary, it can be constructed by interpolation, B-splines or trigonometric functions. Here we use the Rectified Linear Unit (RELU) function for convenience:
\begin{equation}
    \gamma = \operatorname{ReLU}^2\left(\psi_1\psi_2\right) \cdot \operatorname{sgn}(\varphi) ,
    \label{interior}
\end{equation}
The $\operatorname{F}\left(\psi_1,\psi_2\right)$ for the example is shown in \cref{fig_strong_discontinuity} (c) , and the crack embedding is shown in \cref{fig_strong_discontinuity} (d). The introduced function jumps across the crack as well as keeping continuous elsewhere. Besides, the first derivative with respect to the spatial coordinate can be expressed as:
\begin{equation}
    \frac{\partial\gamma}{\partial x_i}=\operatorname{sgn}\left(\varphi\right)\operatorname{ReLU}{\left(\psi_1\psi_2\right)}\sum_{j=1}^{2}\frac{\partial\psi_j}{\partial x_i} ,
    \label{crack_derivative}
\end{equation}
which is also continuous to guarantee the continuity of derivatives in front of the crack tip. 

{\bf Remark .}
    The boundary conditions on crack surface are always intractable for PINNs method based on strong form of governing equations. In most cases, the boundary conditions on crack surface are traction free boundaries.
    Since the neural network solution is separated alongside the crack surface by discontinuity embeddings,
    the nature boundary conditions can be automatically satisfied when the network is trained by energy formulation.


\subsection{Weak discontinuities}

Weak discontinuities occur in heterogeneous solid mechanics problems. 
In these problems, displacements are continuous across the material interface, 
while the normal derivative has a jump discontinuity. 
Solving interface PDE problems with such weak discontinuities has been widely discussed in PINN framework 
\cite{jagtap2020conservative,wu2022inn,he2022mesh,wang2022cenn,wang2020mesh,hu2022discontinuity,tseng2023cusp,yao2023deep,sarma2024interface}.
Here we use a ramp function proposed by Tseng \cite{tseng2023cusp} to produce jump normal derivative:
\begin{equation}
    \gamma_{inter}=\left|\varphi\right| ,
    \label{interface}
\end{equation}
where $\varphi$ is the SDF to track the material interface.
The neural network approximation $NN(\bm{x},\gamma_{inter})$ keeps continuous across the interface. 
Then the normal derivative can be directly derived by the chain rule:
\begin{equation}
    \frac{\partial\bm{u}}{\partial n} = 
    \frac{\partial{NN \left(\bm{x},\gamma_{inter}\right)}}{\partial\bm{x}} \cdot\bm{n} + \frac{\partial{NN\left(\bm{x},\gamma_{inter}\right)}}{\partial\gamma_{inter}}\cdot\frac{\partial\gamma_{inter}}{\partial n} ,
    \label{interface_derivative}
\end{equation}
where ${\partial\gamma_{inter}}/{\partial n}= \operatorname{sgn}(\varphi)$. It can be observed that the jump in normal derivative is allowed:
\begin{equation}
    \llbracket{\frac{\partial\bm{u}}{\partial n}}\rrbracket = 
    \frac{\partial\bm{u}(\bm{x}^+)}{\partial n} - \frac{\partial\bm{u}(\bm{x}^-)}{\partial n} = 
    2 \cdot \frac{\partial{NN(\bm{x}^+)}}{\partial\gamma_{inter}} .
    \label{interface_jump}
\end{equation}

\section{Numerical implementation}\label{Results}

\subsection{Implementation details}\label{details}

    Suitable NN architecture should be constructed for field approximation. The selection of activation function is essential in the NN architecture, as the activation function decides the non-linearity. 
    The ReLU function which is commonly used in deep learning tasks is not suggested in DEM framework, as its first derivative is discontinuous and the second derivative vanishes, which brings difficulties for derive a smooth solution. Here, the tanh function is applied in our implementation for its smoothness. 
    Theoretically, a deeper NN has a stronger expressive ability, but it may suffer from gradient vanishing with tanh activation functions. 
    One technique to mitigate this is the use of residual connections \cite{he2016deep}. The $i-th$ residual block of the NN can be expressed as:
    \begin{equation}
            y^{(i+1)} = y^{(i)} + \operatorname{G}(y^{(i)}, W^{(i)}) ,
        \label{residual_block}
    \end{equation}
    where $\operatorname{G}$ represents the residual mapping with trainable parameters to be learned. In the following cases, two neural networks containing one fully-connected layer and two residual blocks are respectively used for approximating the displacements components $u_1$ and $u_2$ for 2D problems. Each of the residual blocks contains two fully connected layers, and each fully connected layer used in the NN has 30 neurons in all cases.
    
    Then the NN approximation has to fulfill the essential boundary conditions a-priori to guarantee the accuracy with DEM framework \cite{wang2022cenn,nguyen2021parametric}.
    Generally, the essential boundary conditions can be imposed by "hard constraints" on the neural network approximation. Let $\bm{u}(\bm{x}) = NN(\bm{x}, \bm{\gamma}(\bm{x}))$ rewritten to be $\hat{\bm{u}}(\bm{x})$, the hard constraints can be given by the following transformation:
    \begin{equation}
        \bm{u}\left(\bm{x}\right) = 
        A(\bm{x}) \cdot \hat{\bm{u}}(\bm{x})+B(\bm{x}) ,
        \label{boundary_condition}
    \end{equation}
    where $A(\bm{x})$ and $B(\bm{x})$ are constructed to satisfy the given boundary conditions. 
    The hard constraints can be directly constructed for simple cases \cite{lagaris1998artificial}.
    For complex boundaries, it can be decided either by radius basis functions \cite{wang2022cenn}, pre-trained neural networks \cite{sheng2021pfnn} or distance functions \cite{sukumar2022exact}.  

    Finally, the NN is trained to minimize the potential energy, which involves numerical integration for approximating the strain energy. The integration accuracy is essential to the calculation as the integration error will accumulate during the training epochs. In this paper, the composite trapezoidal quadrature rule is implemented. 
    Despite that the trapezoidal rule requires grids in regular domain, it can be implemented on irregular geometries by transforming the domain into a regular reference domain \cite{nguyen2021parametric}. Besides, the sampling grids are also refined in the vicinity of the crack tips due to the higher integration accuracy required by the singular mechanical behavior. During the training process, the Adam algorithm \cite{kingma2017adam} is applied to optimize the trainable parameters. The initial learning rate is set to be 0.02, and will be dropped by 0.5 every 5000 epochs. Besides, the training will be stopped in advance if no improvement is observed in 1000 epochs. The pytorch library \cite{paszke2019pytorch} is used for implementation, and the NNs are trained on one Nvdia RTX 3080ti.

\subsection{A center crack in homogeneous plate} \label{example1}

\begin{figure}[!ht]
        \centering
        \includegraphics[]{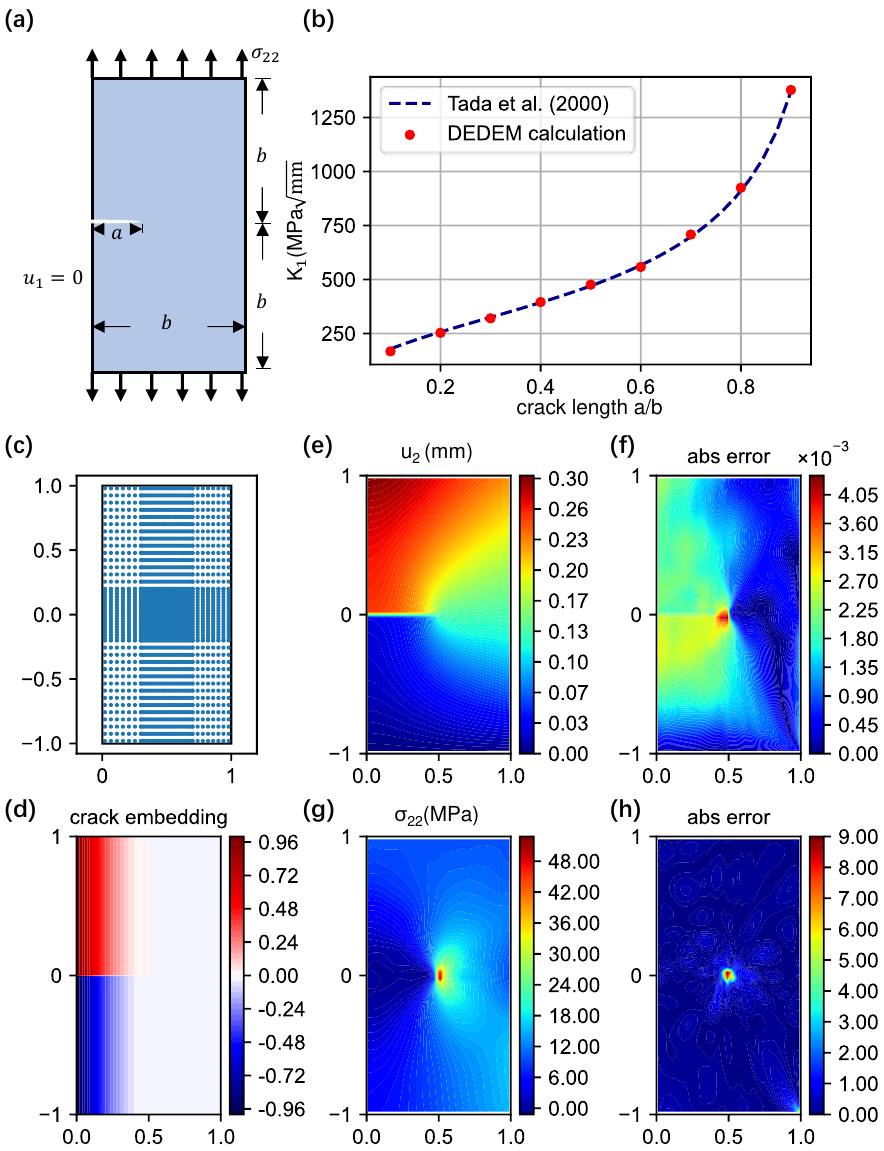}
        \caption{A center crack in a homogeneous plate: 
        (a) the geometry of the symmetric part for computation; 
        (b) Comparison of $K_1$ with the reference solution \cite{Tada2000}; 
        (c) Sampling points for numerical integration and (d) the crack embedding for the case $a/b=0.5$; 
        Predicted solutions and absolute error compared with FEM for the case $a=0.5$m: displacement $u_2$ (e,f) and stress $\sigma_{22}$ (g,h).
        }
        \label{fig_example1}
    \end{figure}

The first example is a center crack in a homogeneous plate under remote tension shown in \cref{fig_example1}(a). The plate is in plane strain condition with the Poisson ratio $\nu$=0.3 and Young's modulus E=100GPa under the tension $\sigma$=10MPa at the top and bottom of the plate. The length and width are taken to be b=1m, and the crack with the length of a is located at y=0. Taking the symmetry into consideration, only the right-hand part is taken for calculation. The left edge is fixed to be $u_1=0$, and the lower edge is fixed to be $u_2=0$. The discontinuity embedding for describing the crack in a half plate can be written as:
    \begin{equation}
        \gamma(\bm{x})=\operatorname{ReLU}^2(a-x_1) \cdot \operatorname{sgn}(x_2).
        \label{embedding_example1}
    \end{equation}
We first take $a=0.5m$ as an example. The value of embedding is plotted in \cref{example1}(d), where discontinuous values can be observed across the crack surface. Then the neural networks are trained by 15000 epochs with 80$\times$100 sampling grid points shown in \cref{fig_example1}(b). The hard constrain for essential boundary conditions is given by:
\begin{equation}
    \begin{aligned}
    & u_1 = \hat{u}_1 \cdot x_1  ,
    & u_2 = \hat{u}_2 \cdot (x_2 + 1) / 2 .
    \label{hardBC_example1}
    \end{aligned}
\end{equation}
To evaluate the results, FEM is performed on a refined mesh. Since the plate is primarily subjected to a mode $\RomanOne$ crack, we focus on the displacement component $u_2$ and the stress component $\sigma_{22}$. The results predicted by DEDEM are presented in \cref{example1}(e, g), with the point-wise absolute errors compared to FEM calculations shown in \cref{example1}(f, h). The results demonstrate that the jump discontinuity in displacement is accurately captured. Additionally, due to the strong non-linearity of the NN, the singular stress field near the crack tip is accurately resolved without requiring any special treatment at the crack tip. 

To further evaluate the proposed method, stress intensity factors (SIFs) are calculated. SIFs are key to predict the failure and propagation of crack. Since the jump discontinuities has been obtained, the SIFs can then be derived by displacement extrapolation method \cite{chan1970finite} with the crack opening displacements  $\delta_i = u_i^+ - u_i^-$ on the crack surface \cite{anderson2005fracture}:
\begin{equation}
\begin{aligned}
    &\delta_1 = K_2 \frac{\kappa+1}{\mu}\sqrt{\frac{r}{2\pi}} + O(r^{3/2}) ,
    &\delta_2 = K_1 \frac{\kappa+1}{\mu}\sqrt{\frac{r}{2\pi}}  + O(r^{3/2}) ,
    \label{homo_COD}
\end{aligned}
\end{equation}
where $\kappa=3-4\nu$ is the kolosov constant for plane strain problems, $\mu_i={E_i}/{2(1+\nu_i)}$ is the shear modulus, and $r$ is the distance of the points on the crack surface to the crack tip. Details of the SIF calculation method can be seen in \cref{appendixA}. We vary the crack length from 0.1m to 0.9m and calculate the corresponding $K_1$ values. For cases where $a/b > 0.2$, points located between $0.3a$ and $0.35a$ behind the crack are selected for the SIF calculation, while points between $0.4a$ and $0.8a$ are chosen for cases where $a/b \leq 0.2$.  The results are compared with the reference solution given by \cite{Tada2000}: 
\begin{equation}
    K_1 = \sigma \sqrt{\pi a} \left[ 1 - 0.025 {\left(\frac{a}{b}\right)}^2 +0.06 {\left(\frac{a}{b}\right)}^4 \right] \sqrt{\sec{\frac{\pi a}{2b}}}.
\end{equation}
A comparison of the calculated $K_1$ values with the analytical solution is given in \cref{fig_example1} (b), with the relative errors listed in \cref{table_example1}. The calculations show excellent agreement with the reference solution, with relative errors no greater than two percent for cases where $a/b>0.1$. The largest relative error occurs for the shortest crack, and a detailed discussion of this phenomenon is provided in \cref{Discussion}.
\begin{table}[!hbtp]
    \begin{center}
    \caption{\label{table_example1} $K_1$ for a center crack in homogeneous plate}     
    \begin{tabular}{cccc}
        \toprule
        a/b & Reference \cite{Tada2000} & Calculation  & Error $(\%)$\\
         & $(\mathrm{MPa\sqrt{mm}})$ & $(\mathrm{MPa\sqrt{mm}})$ & \\
        \midrule
        0.1 &  $178.3$ & $167.1$ & 6.27 \\
        0.2 &  $256.8$ & $253.1$ & 1.43 \\
        0.3 &  $324.7$ & $320.1$ & 1.41 \\
        0.4 &  $393.1$ & $395.3$ & 0.54 \\
        0.5 &  $470.1$ & $475.9$ & 1.22 \\
        0.6 &  $565.6$ & $558.1$ & 1.32 \\
        0.7 &  $697.5$ & $708.1$ & 1.53 \\
        0.8 &  $909.6$ & $924.4$ & 1.63 \\
        0.9 &  $1370.1$ & $1377.1$ & 0.51 \\
        \bottomrule
    \end{tabular}  
    \end{center}
\end{table}

\subsection{A bi-material interface crack}

\begin{figure}[!ht]
    \centering
    \includegraphics[]{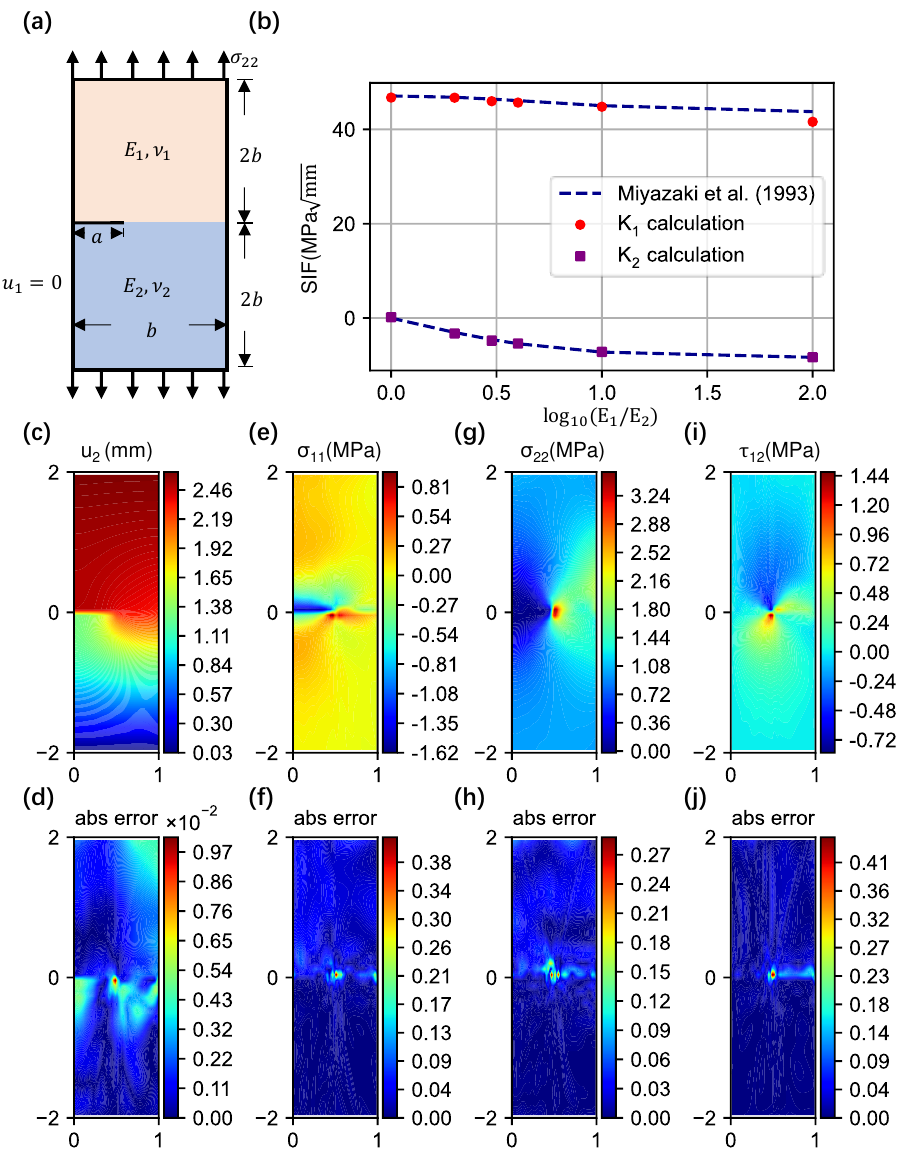}
    \caption{A bi-material interface crack: 
    (a) the geometry of the symmetric part for computation; 
    (b) Comparison of SIFs with the reference solution \cite{miyazaki1993stress}; 
    (c-j): Predicted displacement and stress and absolute error compared with FEM for the case $E_1/E_2=10$.
    }
    \label{fig_example2}
\end{figure}

Analysis for interfacial cracks play an important role in evaluating reliability and mechanical properties for multi-layer structures, as these structures often encounter failure induced by the initiation and propagation of interface cracks resulting from material property mismatches between adjacent layers \cite{Suo19901}. In the second example, we investigate the applicability of DEDEM to bi-material interface cracks. Here we consider a bi-material plate in plane stress condition under the tensile loading $\sigma_{22} = 1$MPa with a center crack at the material interface. Similar to the previous example \cref{example1}, only the right half part is considered and the essential boundary conditions are also same to the previous example \cref{example1}. The geometry is taken to be $b=1$m and the crack length is taken to be $a=0.5$m. Since the normal derivative is discontinuous across the material interface, an embedding is constructed to be input into the neural network to track the weak discontinuity:
\begin{equation}
    \gamma_{\text{interface}}(\bm{x}) = \abs{x_2}.
\end{equation}
Besides, the crack embedding for strong discontinuities across the crack is also required, which is set to be same as the previous example. The Poisson's ratio is set to be $\nu_1=\nu_2=0.3$, and Young's modulus is set to be E$_2=1$GPa, and E$_1=10$GPa. The NN is trained with 100$\times$150 sampling points with 15000 maximum training epochs. Results of DEDEM and absolute errors compared with FEM on displacement and stress are plotted in \cref{fig_example2}(c-j), illustrating high accuracy achieved by the proposed method. It can be observed that the oscillatory crack tip behaviour, which requires complicated crack tip enrichment in classic XFEM \cite{Sukumar20041075} , can be automatically captured without special treatment by the NN solution in the DEDEM framework.
Displacement extrapolation is also applied for computing the SIFs. The elastic mismatch across the material interface results in inseparable tensile and shear effects of interface crack \cite{Rice198898}, making the SIFs cannot be simply decoupled into mode $\RomanOne$ and mode $\RomanTwo$ cracks. The relationship of crack opening displacements and the complex SIF $\mathbf{K} = K_1 + i K_2$ is given by \cite{Hutchinson199163}:
\begin{equation}
    \delta_2+i\delta_1= \frac{8 \textbf{K} r^i\varepsilon \sqrt{r/2\pi} }{(1+2i\varepsilon) \cosh{(\pi\varepsilon)} E^\ast } + O(r^{3/2+i\varepsilon}),
    \label{interface_COD}
\end{equation}
where $i=\sqrt{-1}$ and $\omega$ is a dimensionless bi-material constant defined by the second Dundurs parameter \cite{Dundurs1964650}:
\begin{equation}
    \begin{aligned}
        &\omega=\frac{1}{2\pi} \ln \left(\frac{1-\beta}{1+\beta}\right) ,
        &\beta=\frac{\mu_1(\kappa_2-1)-\mu_2(\kappa_1-1)}
                    {\mu_1(\kappa_2+1)+\mu_2(\kappa_1-1)} ,
        \label{Dundurs}
    \end{aligned}
\end{equation}
where $\mu_m$ and $\kappa_m$ are the shear modulus and Kolosov constant for material $m$ ($m=1,2$), and $\kappa_m=(3-\nu_m)/(1+\nu_m)$ for plane stress problems. We change the E$_1$ from 1GPa to 100GPa, while keeping the E$_2$ to be 1GPa. The computed $K_1$ and $K_2$ are shown in \cref{fig_example2} (b) and \cref{table_example2}. It can be observed that the value of $K_2$ does not equal to 0 even if the plate is under pure tensile loading. Besides, our results agrees well with the calculations of \cite{miyazaki1993stress}.

\begin{table}[!hbtp]
    \begin{center}
    \caption{\label{table_example2} SIFs for a bi-material interface crack.}     
    \begin{tabular}{ccccccc}
        \toprule
        \multirow{3}{*}{E$_1$/E$_2$} & \multicolumn{3}{c}{K$_1$} &  \multicolumn{3}{c}{K$_2$} \\
         \cmidrule(r){2-4} \cmidrule(r){5-7}
         & Reference \cite{miyazaki1993stress} & Calculation  & Error & Reference \cite{miyazaki1993stress} & Calculation  & Error\\
         & $(\mathrm{MPa\sqrt{mm}})$ & $(\mathrm{MPa\sqrt{mm}})$ &  $(\%)$ & $(\mathrm{MPa\sqrt{mm}})$ & $(\mathrm{MPa\sqrt{mm}})$ &  $(\%)$\\
        \midrule
        1 &  $47.08$ & $46.74$ & 0.73 &  $0.00$ & $0.00$ & / \\
        2 &  $46.81$ & $46.72$ & 0.20 &  $-3.05$ & $-3.25$ & 6.72\\
        3 &  $46.41$ & $45.97$ & 0.96 &  $-4.56$ & $-4.79$ & 5.16\\
        4 &  $46.09$ & $45.67$ & 0.91 &  $-5.43$ & $-5.40$ & 0.55\\
        10 &  $45.02$ & $44.80$ & 0.49 &  $-7.21$ & $-7.17$ & 0.65\\
        100 &  $43.76$ & $41.61$ & 4.91 &  $-8.32$ & $-8.31$ & 0.17\\
        \bottomrule
    \end{tabular}  
    \end{center}
\end{table}

\subsection{Multiple cracks} 

\begin{figure}[!ht]
    \centering
    \includegraphics[]{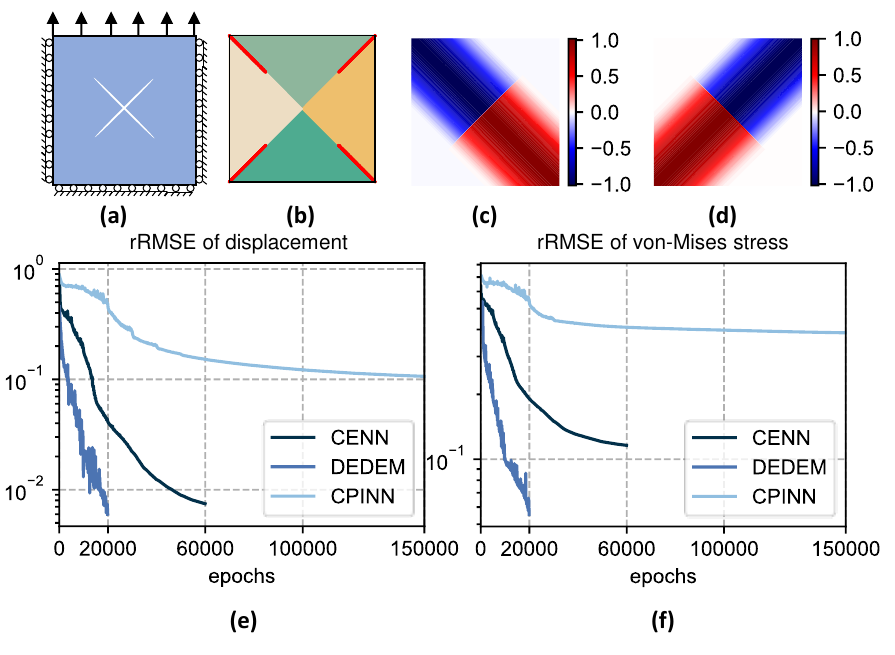}
    \caption{A plate with multiple cracks: 
    (a) The geometry;
    (b) The domain decomposition strategy for CENN and CPINN;
    (c-d) The embeddings to describe cracks for DEDEM;
    The evolution history of rRMSE on (e) displacement and (f) von-Mises stress predicted by CENN, CPINN and DEDEM.
    }
    \label{fig_example3_problem}
\end{figure}

In this section, we aim to investigate the performance of our proposed method on problems with multiple cracks. Here we consider a square plate with a length of 2m shown in \cref{fig_example3_problem} (a), with the center of the plate at the coordinate (0,0). The plate is under plane stress conditions with a uniform tensile load of $\sigma=10$ MPa applied to the top. The bottom of the plate is fixed with $u_2=0$, and the right and left sides are fixed with $u_1=0$. Two cracks intersect at the center of the plate: the first crack runs from the point (-0.5m, -0.5m) to (0.5m, 0.5m), and the second crack runs from (-0.5m, 0.5m) to (0.5m, -0.5m). The Young's modulus is $E=100$ GPa, and the Poisson's ratio is $\nu=0.3$.

To demonstrate the efficiency and accuracy of DEDEM, we compare it with CPINN \cite{jagtap2020conservative} and its energy form, CENN \cite{wang2022cenn}. The domain decomposition strategy for CPINN and CENN is shown in \cref{fig_example3_problem} (b). The entire domain is divided into four subdomains along the interfaces denoted by the red lines ahead of the crack tips. The neural network structure used for each subdomain is the same as in DEDEM but with fewer neurons in each layer (15 neurons for CPINN/CENN and 30 neurons for DEDEM) to maintain a similar total number of trainable parameters. A uniform grid with $N_{pde} = 62500$ internal points is used for training in all three methods. For DEDEM, Two crack embeddings are used for respectively describing the discontinuities across each of the crack, which are depicted in \cref{fig_example3_problem} (c,d) Besides, hard constraints for essential boundary conditions are employed in all the methods:
\begin{equation}
    \begin{aligned}
        &u_1 = \hat{u}_1 \cdot (x_1 + 1) \cdot (x_1 - 1) / 4 ,
        &u_2 = \hat{u}_2 \cdot (x_2 + 1) / 2          .
    \end{aligned}
\end{equation}

The training and optimization strategy for all the methods are same, which is detailed in \cref{details}. The loss of CPINN is given by:
\begin{equation}
    \begin{aligned}
        \operatorname{L}_{\text{CPINN}} &=  \underbrace{\lambda_1 \sum_{i = 1}^{4} \frac{1}{N_{pde}}\sum_{j = 1}^{N_{pde}} {\left\lVert \div{\bm{\sigma}} \right\rVert}^2 }_{\text{equilibrium}}
        + \underbrace{\lambda_2 \sum_{i = 1}^{4} \frac{1}{N_{b}}\sum_{j = 1}^{N_{b}} {\left\lVert \bm{\sigma} \cdot \bm{n} - \bm{\bar{t}} \right\rVert}^2 }_{\text{external boundaries}} 
        + \underbrace{\lambda_3 \sum_{i = 1}^{8} \frac{1}{N_{b}}\sum_{j = 1}^{N_{b}} {\left\lVert \bm{\sigma} \cdot \bm{n} \right\rVert}^2 }_{\text{crack surfaces}}  \\
        &+ \underbrace{\lambda_4 \sum_{i = 1}^{4} \frac{1}{N_{I}}\sum_{j = 1}^{N_{I}} {\left\lVert \bm{u}^{+} - \bm{u}^{-} \right\rVert}^2 
        + \lambda_5 \sum_{i = 1}^{4} \frac{1}{N_{I}}\sum_{j = 1}^{N_{I}} {\left\lVert \left( \bm{\sigma}^{+} - \bm{\sigma}^{-} \right) \cdot \bm{n} \right\rVert}^2 }_{\text{interfaces}}  ,
        \label{CPINN_loss}
    \end{aligned}
\end{equation}
where $N_b$ is the number of sampling points on each boundary, which is taken to be 250. The Neumann boundary conditions on all of the external boundaries and the crack surfaces needs to be taken into consideration, and boundary conditions on each crack surface needs to be considered twice by two subdomains. $N_I$ is the number of points on each interface, which is taken to be 500. $(\cdot)^+$ and $(\cdot)^-$ denotes the solution on different side of the interface predicted by different NNs. The weights of the loss terms $\lambda_i$ is crucial to the training. Although several techniques has been developed \cite{bischof2021multi,wang2022NTK,wang2021understanding,mcclenny2023self,xiang2022self}, it remains to be cumbersome to tune these hyperparameters. In this case, we empirically set $\{\lambda_i\}_{i=1}^5=\{10,1,1,1000,1\}$. For CENN, the loss function is given by:
\begin{equation}
        \operatorname{L}_{\text{CENN}} = J(\bm{u}) + \lambda_I \sum_{i = 1}^{4} \frac{1}{N_{I}}\sum_{j = 1}^{N_{I}} {\left\lVert \bm{u}^{+} - \bm{u}^{-} \right\rVert}^2 ,
\end{equation}
where $J(\bm{u})$ is the potential energy shown in \cref{potential_energy}, and $\lambda_I=-1000\ln{\left(\tanh{\left(N_{I}/N_{domain}\right)}\right)}$ as suggested in \cite{wang2022cenn}. For DEDEM, two separate embeddings are introduced to respectively represent the two cracks, and only $J(\bm{u})$ is required in loss function since no domain decomposition is applied and essential boundary conditions has been imposed. Details of the hyperparameters can be seen in \cref{table_example3}.

\begin{table}[!hbtp]
    \begin{center}
    \caption{\label{table_example3} Hyperparameters and comparison of various methods for the cross crack problem.}     
    \begin{tabular}{ccccc}
        \toprule
        & & CPINN & CENN & DEDEM \\
        \midrule
        \multirow{6}{*}{Hyperparameters} & subdomains & 4 & 4 & / \\
         & trainable parameters & 4328 & 4238 & 4082 \\
         & internal points & 62500 & 62500 & 62500 \\
         & boundary points & 3000 & 250 & 250 \\
         & interface points & 4000 & 4000 & / \\
         & training epochs & 150000 & 60000 & 20000 \\
        \hline
        \multirow{3}{*}{Results} & time (seconds/100 epochs) & 12.41 & 3.94 & 2.35  \\
          &  rRMSE of $\overline{u}$  & 0.107 & 0.008 & 0.0006 \\
                &   rRMSE of $\overline{\sigma}$  & 0.386 & 0.116 & 0.056 \\
        \bottomrule
    \end{tabular}  
    \end{center}
\end{table}

The training time for every 100 epochs are also shown in \cref{table_example3}. Note that the efficiency of all these three methods can be further improved. CPINN and CENN can be accelerated by parallelization techniques \cite{shukla2021parallel} and DEDEM can be accelerated by manually pre-computing the spatial derivative of embeddings. The accuracy of the results are evaluated by the relative rooted mean square error (rRMSE) compared with FEM. The rRMSE of the physical field $\phi$ is given by:
\begin{equation}
        \operatorname{rRMSE}(\phi) = \sqrt{\frac{\int_{\Omega}{ {\left\lVert \phi - \phi_{\text{FEM}} \right\rVert}^2 d\Omega}}
                {\int_{\Omega}{ {\left\lVert \phi_{\text{FEM}} \right\rVert}^2 d\Omega}}}.
\end{equation}

\begin{figure}[!hpt]
    \centering
    \includegraphics[]{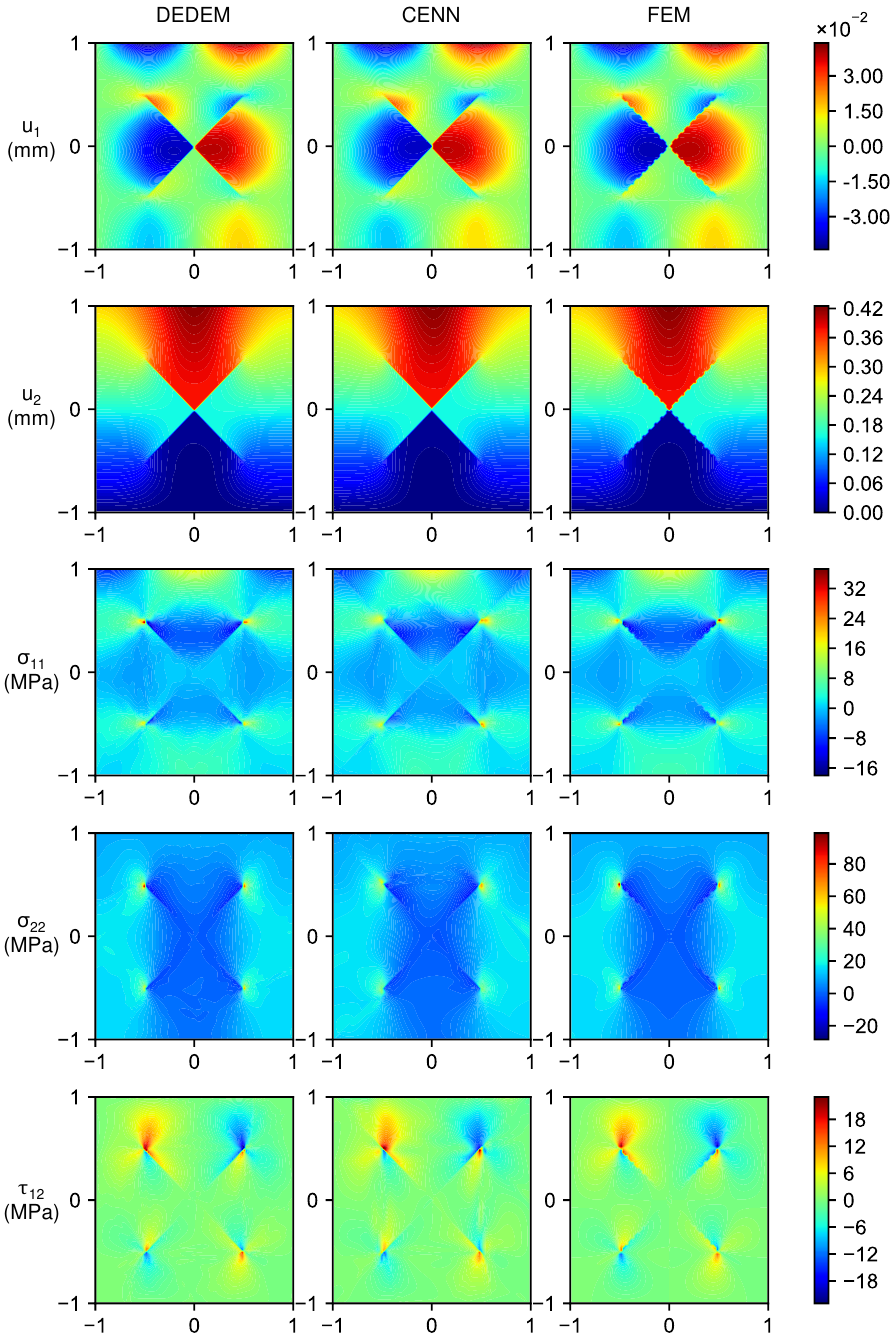}
    \caption{
    Comparison of displacement and stress components of DEDEM, CENN and FEM.
    }
    \label{example3_result}
\end{figure}
To evaluate the accuracy of the three methods, we compare the displacement $\overline{u}=\sqrt{u_1^2+u_2^2}$ and von-Mises stress $\overline{\sigma}=\sqrt{\left( \sigma_{11} - \sigma_{22} \right)^2 / 2 + 3 \sigma_{12}^2}$. The training histories are shown in \cref{example3_result} (e,f), indicating that the DEDEM method achieves best accuracy both on the stress and displacement. Besides, it can be observed that domain decomposition methods require much more training epochs than DEDEM owing to the competition between different loss terms. Detail comparisons of displacement and stress components of DEDEM, CENN and FEM are shown in \cref{example3_result}. The displacement of DEDEM and CENN achieves similar accuracy, while DEDEM performs better on stress prediction, especially in the vicinity of the crack tips.

\subsection{Crack propagation under shear loading} 
\begin{figure}[!htp]
    \centering
    \includegraphics[]{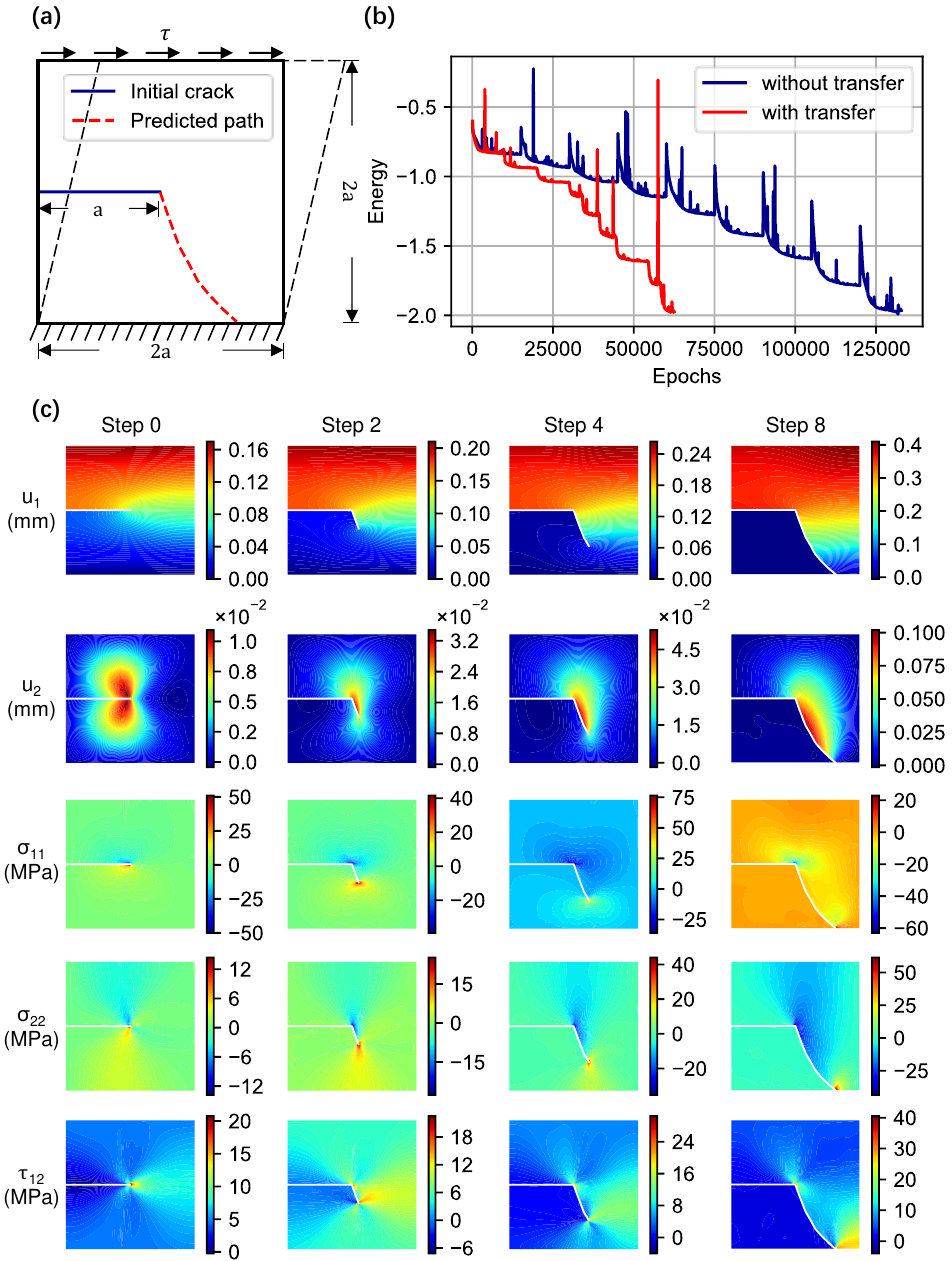}
    \caption{Crack propagation under shear loading: 
    (a) The geometry and the predicted path; 
    (b) Training history with/without transfer learning; 
    (c) Results of displacement and stress components in different steps.
    }
    \label{example4}
\end{figure}

Since SIFs can be calculated from the predicted displacements, the direction of the crack growth can be determined according to the fracture criterion, and the propagation path can then be predicted via iteratively computing the mechanical response of the growing crack. In the last example, we investigate the applicability of DEDEM on predicting the crack path of an edge shear crack. The problem we consider here is shown in \cref{example4} (a). The plate is fixed to be $u_1=0$ on the bottom, and $u_2=0$ on all of the four sides. There is a fixed load $\tau=5$MPa on top of the plate during the propagation process. The plate is in plane stress condition, and the material properties are set to be E=200GPa, $\nu$=0.3.

We assume that the loading process is quasi-static, and no unstable propagation occurs during the process. The maximum circumferential stress hypothesis \cite{Erdogan1963Crack} is applied to determine the propagation direction. Details can be seen in \cref{appendixB}. The growth in crack length is set to be 0.15m, and the embedding to describe the crack will be updated  for each step. A fixed uniform grid with 250$\times$250 sampling points is employed for training during the whole process. The maximum training epochs in each step is set to be 20000. The predicted crack path is shown in \cref{example4} (a). Crack kinking is observed in the first step, then the crack eventually extends to the right part of the bottom side.

Due to the similarities in the mechanical responses of different steps, transfer learning techniques can be applied to accelerate the training process \cite{goswami2020transfer}. The trained NN parameters of the first step is used as the initial parameters for training the current step. The comparison of total training epochs of training with/without transfer learning is shown in \cref{example4} (b), showing that transfer learning effectively reduced the computational cost. It should be noted that the training process can be further accelerated by using the NN parameters in previous step as the initial parameters as more similarities exist in adjacent steps. However, training failure may probably occur due to the error accumulated in previous steps. Therefore, we do not adopt such transfer learning strategy to ensure the robustness. The displacements and stress in different computation steps are shown in \cref{example4} (c), illustrating that our method can successfully simulate the whole cracking process.

\section{Discussion}\label{Discussion}
\subsection{Heterogeneous problem}
\begin{figure}[!htp]
    \centering
    \includegraphics[width=16cm]{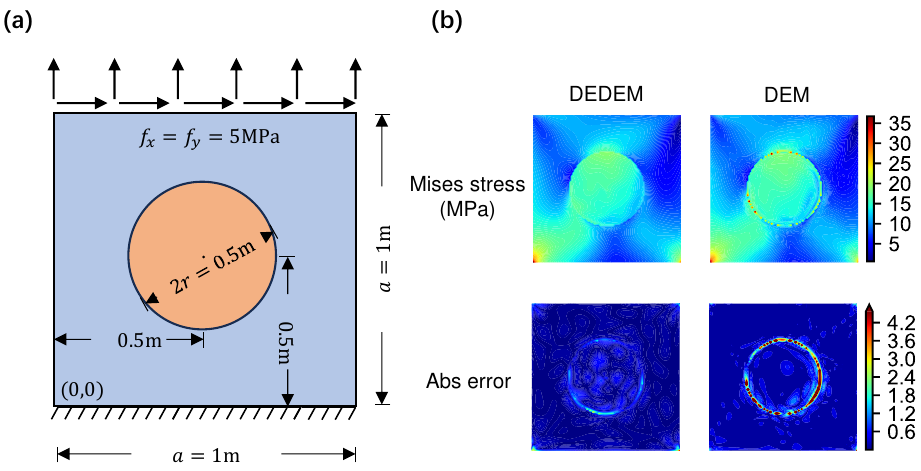}
    \caption{A plate with round inclusion:
    (a) The geometry; 
    (b) Comparison of von-Mises stress predicted by DEDEM and DEM.
    }
    \label{fig_discussion}
\end{figure}

For heterogeneous problems, weak discontinuities occur at the material interfaces between matrix and inclusion, resulting in low accuracy for stress prediction in DEM. Our proposed DEDEM method can improve the accuracy at the material interface without additional computational cost. Here we take a plate with a circle inclusion at the center shown in \cref{fig_discussion}(a) as an example to compare the accuracy of DEDEM and DEM on stress prediction. The elastic property of the center inclusion is $E_{in}=100$GPa; $\nu=0.3$ , while the outer matrix is taken to be $E_{mat}=20$GPa; $\nu=0.3$. The embedding for describing the material interface can be written as:
\begin{equation}
    \gamma(x)=\left|d(x)-r\right|,
\end{equation}
where $d(x)=\sqrt{(x_1-0.5)^2+(x_2-0.5)^2}$ is the distance to the center of the inclusion, and $r=0.25$m is the radius of the inclusion. Both of the methods are trained with 22500 uniformly distributed points. DEDEM converged at around 12000 epochs, and DEM converged at around 20000 epochs. The results of von-Mises stress and the absolute error to FEM calculation are shown in \cref{fig_discussion}(b). It can be observed that DEDEM achieves better accuracy near the material interface.

It should be noted that the low accuracy of DEM near the material interface is caused by the feature of MLP, as MLP with smooth activation functions lacks the ability to fit non-smooth functions. Therefore, if non-smooth trial functions are selected, such as the Kolmogorov-Arnold Networks (KAN) \cite{liu2024kan} constructed by piece-wise splines, weak discontinuities can be automatically learned by NN approximation \cite{wang2024kolmogorov}.

\subsection{Crack-tip singularity}
In many classic numerical methods, such as XFEM and BEM, special singular basis are necessary to model the crack-tip mechanical behavior. However, thanks to the powerful expression ability of NN, singularities can be automatically captured with high accuracy by NN solutions without any special treatment in DEDEM framework as shown in previous examples. It should be pointed out that explicit singularity descriptions can also be incorporated into our proposed method to further improve the efficiency and accuracy. One of the feasible method is to introduce fundamental solutions of the near-tip field to the NN solution \cite{gu2023enriched,gu2024interface,chen2024crack}. However, such method has several shortcomings: (1) It adds trouble for constructing the hard constrain of essential boundary conditions, which is pivotal for training with energy form; (2) The fundamental solution for different crack types, such as homogeneous cracks and interface cracks, are different, which adds complexity for implementation, especially for crack propagation. Another method is to introduce an embedding into the input to impose the singularity feature to the NN solution. However, we found that this approach greatly increases the non-linearity of the NN, which often results in training failure caused by integration errors. In conclusion, despite that it is nonessential to embed crack tip singularity in DEDEM, it remains to be an open problem to find a robust and reliable method to incorporate more priori knowledge of the crack tip mechanical behavior to improve the efficiency.

\subsection{Short cracks}

In \cref{example1}, we found that DEDEM achieves lowest accuracy on samples where the size of the crack is relatively small compared with the geometry. This is because of the following reasons: 1. Short cracks has little influence to the total potential energy, whose feature may be ignored during the training process of minimizing the potential energy; 2. The small crack can be viewed as a local high frequency feature, and it has been reported that NN performs poorly in fitting high frequency components \cite{wang2022NTK}. To this end, our proposed method can be improved by incorporating approaches aiming to high frequency problems, for example, using specific NNs to focus on specific local patterns \cite{moseley2023finite,dolean2024multilevel}.

\subsection{Integration accuracy}
Integration accuracy is vital to the training success. In this paper, dense integration points are set, and local refinement strategy is employed at the crack tips. However, due to the lack of theoretical guidance on the integration accuracy of NN, no strategy can completely guarantee the reliability of the integration, which means that the probability of training failure caused by integration error and gradient pathology cannot be completely eliminated. Similar phenomena have also been reported by some researches for solving crack problems from damage perspective \cite{zheng2022physics}. To avoid the trouble of integration, some researches substitute the automatic differentiation by mesh discretization to compute the derivatives, when the NN is only used to predict the nodal displacements \cite{manav2024phase}. However, such method gives up the mesh-free property of PINNs method. In a word, there still lacks reliable integration scheme within DEM framework.

\section{Conclusion} \label{conclusion}

In this paper, we propose a novel method called discontinuity embedded deep energy method (DEDEM) to solve discontinuous elastic problems via physics-informed machine learning. In this method, the cracks and interfaces are described by discontinuous functions constructed by signed distance functions, and then the NN solutions are imposed to satisfy specific discontinuous features by embedding the discontinuous description to the inputs of NN. The proposed method is demonstrated to be efficient and accurate for modeling the strong discontinuities across the crack, singularity at the crack tip and weak discontinuities across the material interfaces. In addition, DEDEM is much more computationally efficient for solving problems with ideal line cracks compared with the state-of-the-art PINNs methods based on domain decomposition techniques as no additional loss term is required for dealing with the discontinuities; Besides, DEDEM is easy for implementation, and arbitrary discontinuous interfaces and internal boundaries can be tracked and solved. 

The proposed method can be extended in several ways in the future. First, DEDEM can be applied to solve 3D fracture problems; In addition, the proposed method can be easily employed to solve multi-physics fracture problems since the method view the cracks as ideal line cracks and no special treatment is required for the crack tip; Besides, the method can be further developed to solve dynamic fracture; Furthermore, a data-efficient surrogate model can be established for fracture problems by combining the operator learning frameworks \cite{lu2019deeponet,li2020fourier}. In conclusion, DEDEM provides a new idea for solving discontinuous PDEs within PINNs framework. We believe that the proposed method is promising to be applied to forward calculation, inverse modeling and surrogate modeling for a large variety of fracture problems.

\appendix

\section{Calculation of the stress intensity factors}\label{appendixA}

\renewcommand{\theequation}{A.\arabic{equation}}
\setcounter{equation}{0}
\renewcommand{\thefigure}{A.\arabic{figure}}
\setcounter{figure}{0}
\renewcommand{\thetable}{A.\arabic{table}}
\setcounter{table}{0}

Stress intensity factors (SIFs) are crucial parameters to fracture problems for determine the crack propagation and material failure. Here, we use the displacement extrapolation method to derive SIFs. The main idea of this method is to utilize the crack opening displacement (COD) to fit the SIFs. For homogeneous and linear elastic materials, the relationship of COD and SIFs are given by \cref{homo_COD}. The relation can be rewritten as:
\begin{equation}
    \left\{ \begin{aligned}  & {\widetilde{K}}_1(r) \\ & {\widetilde{K}}_2(r) \end{aligned} \right\} =
    \frac{\mu}{\kappa+1}\sqrt{\frac{2\pi}{r}}
    \left\{ \begin{aligned}  & \delta_{2} \\ & \delta_{1} \end{aligned} \right\} = 
    \left\{ \begin{aligned}  & K_{1} \\ & K_{2} \end{aligned} \right\}  + O(r) \approx
    \left\{ \begin{aligned}  & K_{1} \\ & K_{2} \end{aligned} \right\} +
    \left\{ \begin{aligned}  & c_{1} \\ & c_{2} \end{aligned} \right\} r.
\end{equation}
Then linear regression technique is employed to fit ${\widetilde{K}}_i(r)$ $(i=1,2)$, and the SIFs can then be decided by the intercept:
\begin{equation}
    \left\{ \begin{aligned}  & K_{1} \\ & K_{2} \end{aligned} \right\}  = \frac{\mu \sqrt{2\pi}}{\kappa+1} \lim_{r\rightarrow0} \frac{1}{\sqrt{r}}
    \left\{ \begin{aligned}  & \delta_{2} \\ & \delta_{1} \end{aligned} \right\}.
\end{equation}
Similarly, for bi-material interface cracks, the SIFs can be calculated by using \cref{interface_COD}. The real and imaginary parts are separated, and the relation can be rewritten as:
\begin{equation}
    \left\{ \begin{aligned}  & {\widetilde{K}}_1(r) \\ & {\widetilde{K}}_2(r) \end{aligned} \right\} = D
    \begin{bmatrix}  
    -2\varepsilon \cos{Q} + \sin{Q} &  \cos{Q} + 2\varepsilon\sin{Q} \\ \cos{Q} + 2\varepsilon\sin{Q} & 2\varepsilon \cos{Q} - \sin{Q} \end{bmatrix} 
    \left\{ \begin{aligned}  & \delta_{2} \\ & \delta_{1} \end{aligned} \right\} \approx
    \left\{ \begin{aligned}  & K_{1} \\ & K_{2} \end{aligned} \right\} +
    \left\{ \begin{aligned}  & c_{1} \sin{Q} \\ & c_{2} \cos{Q} \end{aligned} \right\} r,
\end{equation}
where
\begin{equation}
    D = \frac{2\mu_1\mu_2\cosh{(\pi\varepsilon)}}{\mu_1(1+\kappa_2)+\mu_2(1+\kappa_1)}\sqrt{\frac{2\pi}{r}} \text{, } Q = \varepsilon\ln{r}.
\end{equation}

\section{Maximum circumferential stress criterion}\label{appendixB}

\renewcommand{\theequation}{B.\arabic{equation}}
\setcounter{equation}{0}
\renewcommand{\thefigure}{B.\arabic{figure}}
\setcounter{figure}{0}
\renewcommand{\thetable}{B.\arabic{table}}
\setcounter{table}{0}

After obtaining the SIFs, the propagation direction of the crack can be determined by fracture criterion. In this paper, the maximum circumferential stress criterion is applied. The asymptotic near-tip circumferential stress takes the following form:
\begin{equation}
    \sigma_{\theta} = \frac{1}{2\sqrt{2\pi r}} \cos{\frac{\theta}{2}}\left[ K_1 (1+\cos{\theta}) - 3 K_2 \sin{\theta}\right],
\end{equation}
where $r$ and $\theta$ are the local polar coordinates to the crack tip. The criterion assume that the crack will propagate from its tip in a direction $\theta_c$ where the $\sigma_{\theta}$ reaches maximum. The maximum circumferential stress satisfies the following expression:
\begin{equation}
    \pd{\sigma_{\theta}}{\theta} = \frac{-3}{4\sqrt{2\pi r}} \cos{\frac{\theta}{2}}\left[ K_1 \sin{\theta} + K_2 (3\cos{\theta}-1)\right] =0.
\end{equation}
Then $\theta_c$ can be given by:
\begin{equation}
    K_1 \sin{\theta_c} + K_2 (3\cos{\theta_c}-1) =0.
\end{equation}
Solving this equation gives:
\begin{equation}
   \theta_c = \arccos{\frac{3 K_2^2 \pm \sqrt{K_1^4 + 8 K_1^2 K_2^2}}{K_1^2 + 9 K_2^2}}.
\end{equation}

%
\bibliographystyle{elsarticle-num}
\bibliography{scopus}
\end{document}